\documentclass[twocolumn,tighten]{aastex63}
\usepackage{needspace}
\usepackage{tabularx}
\usepackage{booktabs}
\usepackage{multirow}
\usepackage{tablefootnote}
\let\tablenum\relax % Let siunitx redefine tablenum
\usepackage{siunitx}

\newcommand{\tess}{\textit{TESS}}

\newcommand{\Gaia}{\textit{Gaia} DR3}

\newcommand\Msun{$M_{\odot}$}
\newcommand\Rsun{$R_{\odot}$}
\newcommand\Lsun{$L_{\odot}$}
\newcommand\RE{$R_{\Earth}$}
\newcommand\ME{$M_{\Earth}$}

\def\vplanet{\texttt{{VPLanet}}}
\def\atmesc{\texttt{{AtmEsc}}}
\def\stellar{\texttt{{STELLAR}}}

\newcommand{\name}{TOI-4364}
\newcommand{\pname}{\texorpdfstring{TOI-4364\,b}{TOI-4364\,b}} % hyperref complains that the PDF bookmark cannot contain the "\," space. The second argument to \texorpdfstring is the alternative text that latex should write to the PDF bookmark
\newcommand{\mass}{$0.487^{+0.017}_{-0.016}$}
\newcommand{\age}{$710 \pm 100$}

\newcommand{\protne}{$10.57$}
\newcommand{\gaiamag}{$12.534 \pm 0.006$} %gaia dr3 
\newcommand{\dist}{$43.85 \pm 0.04$} %gaia dr3
\newcommand{\rad}{$0.4684^{+0.0087}_{-0.010}$}
\newcommand{\prad}{$2.01^{+0.10}_{-0.08}$\,\RE{}}
\newcommand{\teq}{$488^{+4}_{-4}$}

\newcommand{\feh}{$0.148 \pm {0.046}$}
\newcommand{\pmass}{$5.3^{+1.9}_{-1.2}$\,\ME{}} 
\newcommand{\tsm}{44.2}
\newcommand{\esm}{4.9}
\shortauthors{Distler et al.}

\begin{document}
\title{\tess{} Hunt for Young and Maturing Exoplanets (THYME) XII: A Young Mini-Neptune on the Upper Edge of the Radius Valley in the Hyades Cluster}
\author[0009-0006-4294-6760]{Adam Distler}
\email{ajdistler@wisc.edu}
\affiliation{Department of Astronomy,  University of Wisconsin-Madison, 475 N.~Charter St., Madison, WI 53706, USA}

\author[0000-0001-7493-7419]{Melinda Soares-Furtado}
\altaffiliation{NASA Hubble Postdoctoral Fellow}
\affiliation{Department of Astronomy,  University of Wisconsin-Madison, 475 N.~Charter St., Madison, WI 53706, USA}
\affiliation{Department of Physics and Kavli Institute for Astrophysics and Space Research, Massachusetts Institute of Technology, Cambridge, MA 02139, USA}

\author[0000-0001-7246-5438]{Andrew Vanderburg}
\affiliation{Department of Physics and Kavli Institute for Astrophysics and Space Research, Massachusetts Institute of Technology, Cambridge, MA 02139, USA}

\author[0000-0002-7382-0160]{Jack Schulte}
\affiliation{Center for Data Intensive and Time Domain Astronomy, Department of Physics and Astronomy, Michigan State University, East Lansing, MI 48824, USA}

\author[0000-0002-7733-4522]{Juliette Becker}
\affiliation{Department of Astronomy,  University of Wisconsin-Madison, 475 N.~Charter St., Madison, WI 53706, USA}

\author[0000-0003-3654-1602]{Andrew W.~Mann}
\affiliation{Department of Physics \& Astronomy, The University of North Carolina at Chapel Hill, Chapel Hill, NC 27599-3255, USA}

\author[0000-0002-2532-2853]{Steve B.~Howell}
\affiliation{NASA Ames Research Center, Moffett Field, CA 94035, USA}

\author[0000-0001-9811-568X]{Adam L.~Kraus}
\affiliation{Department of Astronomy, The University of Texas at Austin, Austin, TX 78712, USA}

\author[0000-0003-1464-9276]{Khalid Barkaoui}
%- Email: khalid.barkaoui@uliege.be 
\affiliation{Astrobiology Research Unit, Universit\'e de Li\`ege, All\'ee du 6 Ao\^ut 19C, B-4000 Li\`ege, Belgium}
\affiliation{Department of Earth, Atmospheric and Planetary Sciences, Massachusetts Institute of Technology, Cambridge, MA 02139, USA}
\affiliation{Instituto de Astrof\'isica de Canarias (IAC), Calle V\'ia L\'actea s/n, 38200, La Laguna, Tenerife, Spain}

%\author[0000-0001-6637-5401]{Allyson Bieryla}
%\affiliation{Center for Astrophysics \textbar \ Harvard \& Smithsonian, 60 Garden Street, Cambridge, MA 02138, USA}

\author[0000-0001-7124-4094]{C\'{e}sar Brice\~{n}o}
\affiliation{Cerro Tololo Inter-American Observatory, Casilla 603, La Serena, Chile}

%karen.collins@cfa.harvard.edu  (LCO scheduling, data reduction, time contribution)
\author[0000-0001-6588-9574]{Karen A.\ Collins}
\affiliation{Center for Astrophysics \textbar \ Harvard \& Smithsonian, 60 Garden Street, Cambridge, MA 02138, USA}

\author[0000-0003-2239-0567]{Dennis Conti}
\affiliation{American Association of Variable Star Observers (AAVSO), 185 Alewife Brook Parkway, Suite 410, Cambridge, MA  02138 USA}

\author[0000-0002-4715-9460]{Jon M.~Jenkins}
\affiliation{NASA Ames Research Center, Moffett Field, CA 94035, USA}

%\author{Nicholas Law}
%\affiliation{Department of Physics \& Astronomy, The University of North Carolina at Chapel Hill, Chapel Hill, NC 27599-3255, USA}

\author[0000-0002-9521-9798]{Mary Anne Limbach}
\affiliation{Department of Astronomy, University of Michigan, Ann Arbor, MI 48109, USA}

\author[0000-0002-8964-8377]{Samuel~N.~Quinn}
\affiliation{Center for Astrophysics \textbar \ Harvard \& Smithsonian, 60 Garden Street, Cambridge, MA 02138, USA}

\author[0000-0001-8227-1020]{Richard P.~Schwarz}
\affiliation{Center for Astrophysics \textbar \ Harvard \& Smithsonian, 60 Garden Street, Cambridge, MA 02138, USA}

\author[0000-0002-6892-6948]{Sara Seager}
\affiliation{Department of Physics and Kavli Institute for Astrophysics and Space Research, Massachusetts Institute of Technology, Cambridge, MA 02139, USA}
\affiliation{Department of Earth, Atmospheric and Planetary Sciences, Massachusetts Institute of Technology, Cambridge, MA 02139, USA}
\affiliation{Department of Aeronautics and Astronautics, MIT, 77 Massachusetts Avenue, Cambridge, MA 02139, USA}

\author[0000-0001-7836-1787]{Jake D.~Turner}
\affiliation{Department of Astronomy and Carl Sagan Institute, Cornell University, Ithaca, NY, USA}

\author[0000-0002-6778-7552]{Joseph D.~Twicken}
\affiliation{SETI Institute, Mountain View, CA  94043, USA}
\affiliation{NASA Ames Research Center, Moffett Field, CA  94035, USA}

\author[0000-0002-4265-047X]{Joshua N.\ Winn}
\affiliation{Department of Astrophysical Sciences, Princeton University, Princeton, NJ 08544, USA}

\author[0000-0002-0619-7639]{Carl Ziegler}
\affiliation{Department of Physics, Engineering and Astronomy, Stephen F. Austin State University, 1936 North St, Nacogdoches, TX 75962, USA}

\begin{abstract}
We present the discovery and characterization of \pname{}, a young mini-Neptune in the tidal tails of the Hyades cluster, identified through \tess{} transit observations and ground-based follow-up photometry. 
The planet orbits a bright M dwarf ($K=9.1$\,mag) at a distance of 44\,pc, with an orbital period of 5.42\,days and an equilibrium temperature of \teq{}\,K. 
The host star's well-constrained age of 710\,Myr makes \pname{} an exceptional target for studying early planetary evolution around low-mass stars.
We determined a planetary radius of \prad{}, indicating that this planet is
situated near the upper edge of the radius valley. This suggests that the planet retains a modest H/He envelope. As a result, \pname{} provides a unique opportunity to explore the transition between rocky super-Earths and gas-rich mini-Neptunes at the early stages of evolution. Its radius, which may still evolve as a result of ongoing atmospheric cooling, contraction, and photoevaporation, further enhances its significance for understanding planetary development.
Furthermore, \pname{}’s moderately high Transmission Spectroscopy Metric of \tsm{} positions it as a viable candidate for atmospheric characterization with instruments such as JWST. This target has the potential to offer crucial insights into atmospheric retention and loss in young planetary systems.
\end{abstract}

\keywords{exoplanets, star clusters, planets and satellites: individual (\name{})}

\section{Introduction}\label{sec:intro}
Investigations of young exoplanets offer crucial insights into the rapid evolutionary changes that occur within the first few hundred million years after the formation of a planet. 
During this early period, planets undergo significant transformations in their orbital dynamics \citep[e.g.,][]{1998Icar..136..304C, 2006ApJ...649.1004A, 2012AREPS..40..251M}, structural properties \citep{Kite2009}, as well as atmospheric composition and structure \citep[e.g.,][]{Owen2017,Dorn2018,Ginzburg2018}.

To investigate these early evolutionary processes, the \tess{} Hunt for Young and Maturing Exoplanets (THYME; \citealt{Newton2019}) collaboration focuses on detecting and characterizing transiting exoplanets within young stellar groups and open clusters. These environments are advantageous because the ages, distances, and metallicities of the stars are well-constrained, enabling more accurate exoplanetary and host star parameters. 
Consequently, the young exoplanets identified by THYME and similar surveys serve as key benchmarks for testing theories of exoplanet formation, migration, and evolution.

The nearby Hyades cluster, located at a distance of 46\,pc and estimated to be $\approx 700$\,Myr old \citep{Brandner2023}, is an ideal target for this search. 
In this work, we present the discovery, characterization, and follow-up observations of a planet in the tidal tails of the Hyades cluster. This planet, \pname{}, is the eighth detected in the cluster, joining K2-25\,b \citep{Mann2016}, K2-136\,b/c/d \citep{Ciardi2018, Mann2018}, HD\,283869\,b \citep{Vanderburg2018}, HD\,285507\,b \citep{Quinn2014}, and HD\,28305\,b \citep{Sato2007}. 
Although membership in the Hyades has been questioned for HD\,28305 \citep{Tabernero2012, Roser2019} and HD\,283869 \citep{Douglas2014, Oh2020, Long2023}, we include these stars as Hyades members in the figures presented in this work.

\pname{} is found in this work to be a transiting mini-Neptune with a 5.42-day orbit around an M dwarf. With a radius of \prad{}, this planet resides at the upper edge of the radius valley \citep{2017AJ....154..109F}, a region where planets with radii between $1.5-2.0$\,\RE{} are scarce. Typically, planets smaller than the lower bound are rocky, while those larger than the upper bound are gaseous mini-Neptunes. 
Some of the most widely accepted mechanisms proposed to drive the formation of the radius valley are photoevaporation \citep{Owen2017,Burn2024} and core-powered mass loss \citep{Ginzburg2018}.

The youth and $K$-band brightness ($K=9.1$\,mag) of this newly discovered planetary system make it a compelling target for follow-up observations. 
As a young mini-Neptune on the upper edge of the radius valley, \pname{} provides a unique opportunity to explore atmospheric properties and the processes shaping planetary evolution, with its bright, young M dwarf host offering valuable insights into atmospheric retention and loss during the early stages of planetary formation.

The manuscript is structured as follows: \S \ref{sec:data} details the data products used for the detection and characterization of this system; \S \ref{sec:analysis} focuses on the characterization of stellar properties of \name{}, along with the analysis of \pname{} and its orbit; \S \ref{sec:Vetting} examines potential false positive scenarios, \S \ref{sec:discussion} discusses \pname{} in the context of other Hyades planets and within the broader census of exoplanet discoveries.

\section{Data}
\label{sec:data}
To investigate potential planetary transits of the Hyades star \name{}, we used photometric data from the Transiting Exoplanet Survey Satellite (TESS; \citealt{Ricker2015}), photometry and radial velocity (RV) measurements from ground-based facilities, and high-resolution speckle imaging.

\subsection{The TOI-4364 \tess{} Light Curve}
\label{subsec:tess}
The \tess{} survey divides the sky into rectangular sectors. Each sector covers a field of view of $24^{\circ} \times 96^{\circ}$ over a baseline of approximately 26 days. 
To date, \name{} (TIC 4070275) has been observed in Sector 5 and Sector 32 (start dates: 11/11/2018 \& 11/20/2020). 
The data are available with a time sampling of 120 seconds.
Table~\ref{tab:tessphot} lists the relevant sectors, cadence, camera number, CCD number, and date range corresponding to these observations.
\begin{figure*}[tbh!]
    \centering
    \includegraphics[width=0.98\textwidth]{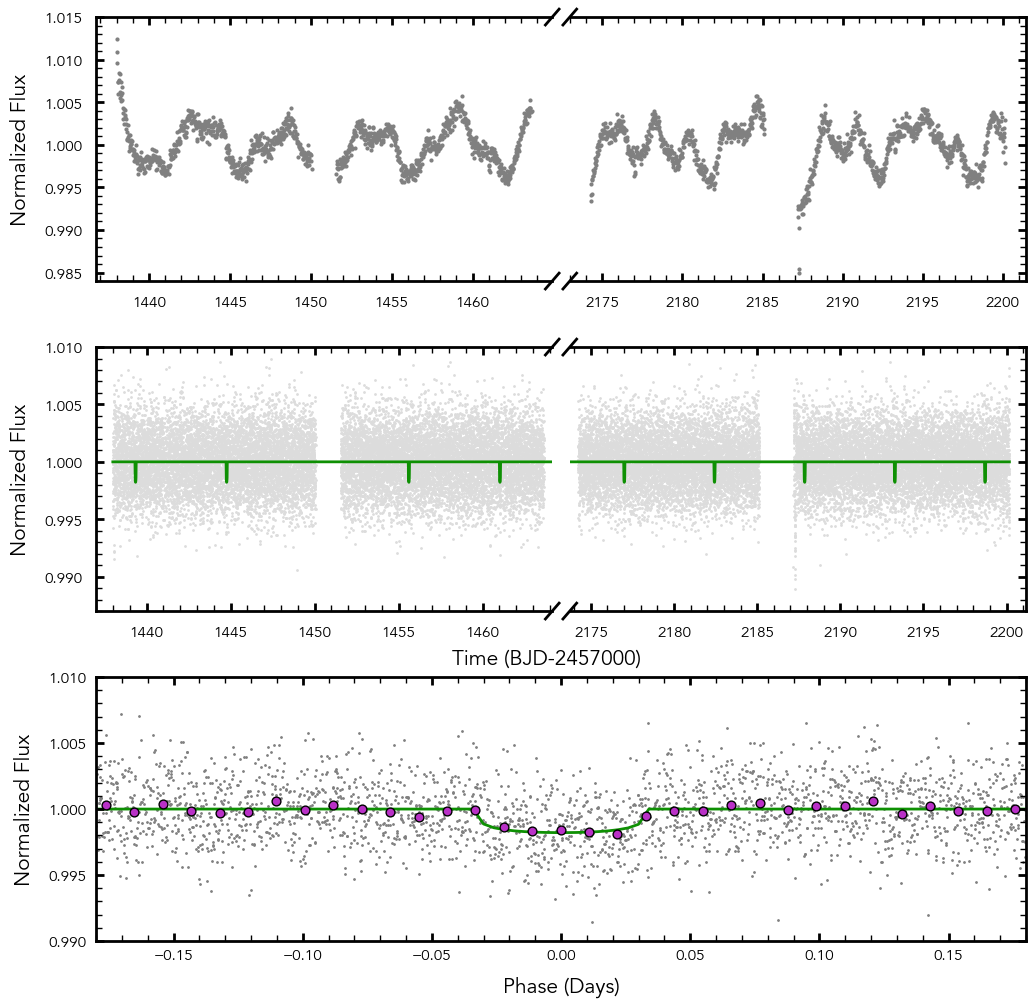}
    \caption{\tess{} light curve analysis for \pname{}. \textbf{Top:} unflattened, processed light curve of \name{}. \textbf{Middle:} flattened, systematically-corrected light curve (grey) with the best fit \texttt{batman} (green). \textbf{Bottom:} phase-folded light curve with binned points (purple) and the phase-folded best-fit transit model (green).}
   \label{fig:MCMCtransitfit}
\end{figure*}

\begin{deluxetable}{ccccc} % Define column alignment
\tablecaption{\tess{} Photometry Used in the Detection and Characterization of \pname{}.} % Table caption
\centering
\label{tab:tessphot} % Label for referencing
\startdata
\\ \multicolumn{5}{c}{\tess{} Data Overview} \\ \hline \hline
Sector & Cadence & Camera & CCD & Observation Dates \\ \hline
5 & 120\,s & 1 & 2 & 2018 Nov 11 – Dec 06 \\
32 & 120\,s & 1 & 2 & 2020 Nov 20 – Dec 15 \\
\hline
\enddata
\end{deluxetable}

The data from both sectors were reduced and analyzed by the Science Processing Operations Center (SPOC) at NASA Ames Research Center \citep{jenkinsSPOC2016}. 
On May 27, 2021, SPOC conducted a transit search of the combined light curve from the two sectors, using an adaptive, noise-compensating matched filter \citep{2002ApJ...575..493J,2010SPIE.7740E..0DJ, 2020ksci.rept....9J}.
This resulted in a Threshold Crossing Event with 5.42\,d period. 
An initial limb-darkened transit model was fitted \citep{2019PASP..131b4506L} and a suite of diagnostic tests were performed to help assess the planetary nature of the signal \citep{2018PASP..130f4502T}. 
The transit signature passed all diagnostic tests presented in the SPOC Data Validation reports, and the source of the transit signal was localized within $1.23 \pm 3.61$" of the target star. The TESS Science Office (TSO) reviewed the vetting information and issued an alert for TOI 4364.01 on August 5, 2021. 

For young, active stars, specialized processing can significantly mitigate \tess{} light curve noise. 
We performed additional systematic correction to the SPOC simple aperture photometry light curve \citep{2010SPIE.7740E..23T, 2020ksci.rept....6M} using the method outlined by \cite{Vanderburg2019}. 
This approach involved simultaneously fitting a basis spline with 0.3-day intervals to model the low-frequency stellar variability, along with several vectors describing spacecraft systematics. These vectors included the mean and standard deviation of the quaternion time series to correct for pointing jitter.

We also incorporated cotrending vectors from SPOC’s band 3 correction to address fast timescale variations \citep{Smith2012,Stumpe2012, Stumpe2014} and included a 0.1-d high-pass filtered version of the background flux time series in the decorrelation. We performed a linear least-squares fit to the light curve and removed the contributions of the systematics vectors to produce a cleaned light curve. 
We then removed the high-frequency stellar variability by subtracting a basis spline fit at 0.5-day breakpoints.

Following these corrections, we searched the light curve for box-shaped dips in brightness using the box-least-squares (BLS) algorithm \citep{2002A&A...391..369K}, investigating a period range of $0.4-26$ days. 
A compelling 5.42-day signal was apparent at a statistical significance of 11.95, calculated following the method of \cite{Kipping2023}. The processed data along with a phase-folded light curve are shown in Figure~\ref{fig:MCMCtransitfit}.

In addition, we explored the existing transit signals using the Transit Least Squares\footnote{\url{https://transitleastsquares.readthedocs.io}} (TLS) algorithm \citep{HippkeTLS2019}.
By incorporating considerations such as limb darkening effects with transit ingress and egress, the TLS algorithm can more accurately model transit shapes, leading to improved detection efficiency. Using the TLS algorithm, we again recovered the transit signal with the same ephemeris.
We attribute the observed periodic signal to the detection of a new planet candidate, \pname{}, with a transit depth of approximately 2\,ppt.
This is consistent with the transit depth expected from a mini-Neptune companion. 

\subsection{Ground-based Seeing-Limited Photometry}
\label{sec:lcogt}

\begin{figure}[htb!]
    \centering
    \includegraphics[width=0.45\textwidth]{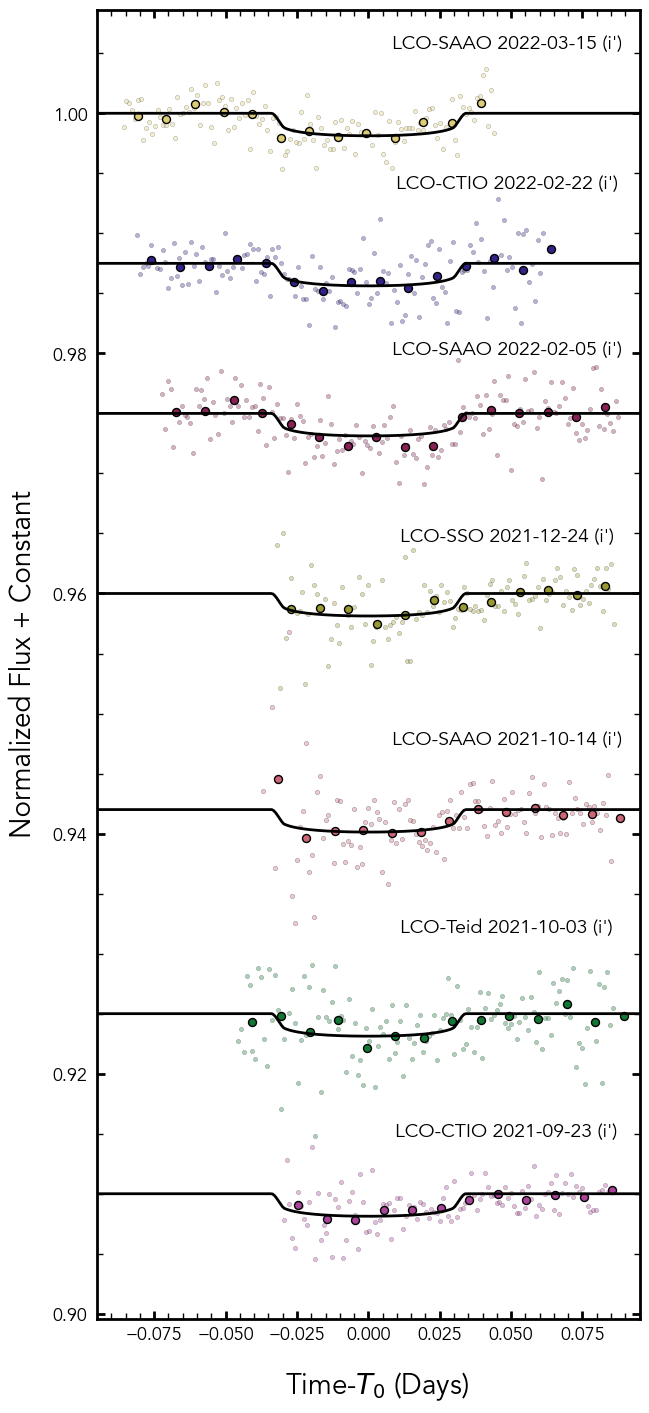}
    \caption{LCOGT 1-m \pname{} follow-up observations (small points) along with binned data (larger circles) and best-fit transit models (black line). Individual observations are shifted along the y-axis for clarity.}
   \label{fig:FUtransitfit}
\end{figure}

Follow-up observations using instruments with smaller pixel plate scales can help determine the source of a potential planetary signal.
This is important given the large pixel plate scale of \tess{}.
Moreover, planetary transit parameters can be refined by reducing third-light contamination (see \citealt{Ciardi2015}) and the ephemeris can be more accurately constrained. 
Follow-up photometric observations were conducted using several 1.0-meter telescopes from the Las Cumbres Observatory Global Telescope (LCOGT; \cite{Brown2013}) Network --- a worldwide array of robotic telescopes designed for continuous sky monitoring. Each telescope is equipped with Sinistro cameras, optimized for high-precision photometry. Observations were made at four sites: the South African Astronomical Observatory (SAAO), the Cerro Tololo Inter-American Observatory (CTIO), the Siding Spring Observatory (SSO), and the Teide Observatory.
The details of these observations are summarized in Table~\ref{tab:fuobs}.

Over six months, seven observing sessions were carried out, with six capturing the full transit and one focusing on the planet's egress. All observations were performed using the \textit{i'} filter, ensuring consistent data quality across sites. 
The science data reduction was performed using the standard LCOGT {\tt BANZAI} pipeline \citep{McCully_2018SPIE10707E}.  Photometric measurements were extracted using uncontaminated apertures, ranging between 4.0-6.6", using the {\tt AstroImageJ}\footnote{\url{https://www.astro.louisville.edu/software/astroimagej}} software \citep{Collins_2017}. These small apertures were chosen to exclude contamination from a nearby (${\sim}11\arcsec$) neighbor that was present in the SPOC apertures. 
\textit{Gaia} DR3 identified no contaminating sources brighter than $G_{\mathrm{Gaia}}>21$\,mag closer than this neighbor. The transits were detected convincingly in six of the seven follow-up observations taken, with the data shown in Figure~\ref{fig:FUtransitfit}.
%The observed transit light curves are presented in Figure~\ref{fig:FUtransitfit}.

\begin{deluxetable}{lcccc} % Define column alignment
\tablecaption{Follow-up ground-based photometry of \pname{}.} % Table caption
\centering
\label{tab:fuobs} % Label for referencing
\startdata
\\ \multicolumn{5}{c}{Las Cumbres Observatory Global Telescope Network} \\ \hline \hline
Location & Duration & Coverage & Aperture & Date \\ \hline
SAAO\tablenotemark{a} & 191\,min & Full & 5.4" &  03/15/2022 \\
CTIO\tablenotemark{b} & 208\,min & Full & 5.1" & 02/22/2022 \\
SAAO & 234\,min & Full & 5.1" & 02/05/2022 \\
SSO\tablenotemark{c} & 174\,min & Full & 5.1" & 12/24/2021 \\
SAAO & 177\,min & Full & 4.0" & 10/14/2021 \\
Teide & 189\,min & Full & 5.0" & 10/03/2021 \\
CTIO & 165\,min & Egress & 6.6" & 09/23/2021 \\
\enddata
\tablenotetext{a}{South African Astronomical Observatory}
\tablenotetext{b}{Cerro Tololo Inter-American Observatory}
\tablenotetext{c}{Siding Spring Observatory}
\end{deluxetable}

\subsection{Radial Velocity Measurements}
\label{sec:recon spec}
Several ground-based RV measurements have been collected for our target, along with \Gaia{} spectra \citep{GaiaCollaboration2021}.
The ground-based spectroscopic follow-up consists of four measurements with the Echelle SPectrograph for Rocky Exoplanets and Stable Spectroscopic Observations (\textit{ESPRESSO}) \citep{2021A&A...645A..96P}, which span over three months.
It also includes two measurements with the Tillinghast Reflector Echelle Spectrograph (\textit{TRES}; \citealt{Szentgyorgyi2007,Furesz08}), which are separated by three days.% one with CHIRON \citep{2013PASP..125.1336T}, 

The \textit{ESPRESSO} data were reduced with the Data Reduction Software (DRS) pipeline, which cross-correlates the observed spectrum with a stellar binary mask, and then fits the cross-correlation output with a Guassian to derive the RV. The \textit{TRES} data were first reduced following the method in \cite{Buchhave2010}.
A CCF multi-order analysis was then applied to the data to best constrain the RV across the entire spectrum. 
The precise and stable measurements from \textit{TRES} and \textit{ESPRESSO} help to distinguish between planetary signals and potential perturbations caused by unseen stellar companions, with all measurements taken at or near quadrature. 
We see that the standard deviation of the \textit{ESPRESSO} data is 9.0\,m\,s$^{-1}$, and the \textit{TRES} data are separated by 78\,m\,s$^{-1}$. When we convert the measurements from the two spectrographs to the standard IAU velocity scale, we find that they match at the 220\,m\,s$^{-1}$ level. All of this is consistent with \name{} not being a short-period radial velocity variable.
The details of these observations are listed in Table~\ref{tab:radvels}. 

\begin{deluxetable}{lcccc}[htb!]
\label{sec:RVs}
\centering
%\LARGE
%\tabletypesize{\scriptsize}
\tablewidth{0pt}
\caption{Radial Velocity Measurements of \name{}.
\label{tab:radvels}}
\tablehead{\colhead{Name} & \colhead{RV} & \colhead{err} & \colhead{Exp.~time}& \colhead{Time} \\ %\colhead{Date(s)}
 & \colhead{($km\,s^{-1}$)} & \colhead{($km\,s^{-1}$)} & \colhead{($s$)}& }
\startdata
%Gaia DR2 & 45.98 & 1.12 &  2015.5\\ \cline{1-4}
\Gaia{}\tablenotemark{a} & 44.77 & 0.46 &-- & 2016.0 \\ 
\textit{TRES} & 44.969 & 0.104 & 2160&  2459522.907 \\%11/4/2021 \\
TRES & 44.980 & 0.104 &2400 & 2459525.825 \\ %11/7/2021 \\
%CHIRON & RV & err & Date \\ \cline{1-4}
ESPRESSO & 45.087 & 0.004 & 1800 & 2459795.889 \\ %8/4/2022 \\ 
ESPRESSO & 45.108 & 0.001 &1800 & 2459812.881 \\ %8/21/2022 \\
ESPRESSO & 45.103 & 0.003 &1800 & 2459813.828 \\%8/22/2022 \\
ESPRESSO & 45.109 & 0.002 &1800 & 2459893.824 \\ %11/10/2022 \\ 
\enddata
\tablenotetext{a}{\Gaia{} RV measurements are averaged over multiple visits over 34 months with an average of 8 transits per star.}
\end{deluxetable}

\subsection{High Resolution Speckle Imaging}
To investigate potential contamination from stars at small angular separations (possible variables, eclipsing binaries, or bound companions), we performed high-resolution speckle imaging. 
By analyzing the speckle data, we can investigate whether the observed transits are influenced by additional ``third-light” from close stellar companions. 
This ensures that the planetary signal remains free from contamination by additional stellar light, which is crucial for accurate characterization and validation of the planet \citep{Furlan2017, Furlan2020}.

The first Gemini speckle observation for \name{} occurred on October 16, 2021, and was made with the 'Alopeke instrument on Gemini North in Hawaii. The second Gemini observation took place on Gemini South in Chile on January 5, 2023, using the Zorro instrument under conditions of better seeing and offering more time on target. 
The two instruments are identical, using Electron Multiplying Charge-Coupled Devices (EMCCDs) to enable fast readout and provide simultaneous two-color imaging \citep{Scott2021}. Both observations of \name{} were performed using narrow band filters at two separate wavelengths, 832/40\,nm and 562/54\,nm. This dual-wavelength approach provides a more comprehensive assessment of any detected nearby sources.  The speckle data was reduced using our standard pipeline described in \cite{Howell2011}. No sources were detected to our magnitude limits, as shown in the top panel of Figure~\ref{fig:contrast}.

On October 18, 2021, a broader search for stellar companions was conducted using speckle imaging with the 4.1-meter Southern Astrophysical Research (SOAR) telescope \citep{Tokovinin2018}. The observation was made in the Cousins $I$-band (879 (289)\,nm), a visible bandpass similar to \textit{TESS}. Further details about these observations can be found in the SOAR \textit{TESS} survey \citep{Ziegler2020}. No nearby stars were detected within 3\arcsec of TOI-4364 in the SOAR observations, as shown in the bottom panel of Figure~\ref{fig:contrast}.
Beyond the outer limits of these observations, the \Gaia{} catalog establishes that there are no neighbors brighter than $G_{\mathrm{Gaia}}  \approx 21$\,mag and within 10".
Further, a search of \Gaia{} finds that there are no very wide co-moving, co-distant companions brighter than this magnitude limit out to a projected separation of 1200" (50,000\,AU). 
Therefore, the system does not host any wide binary companions that might induce long-term dynamical evolution of the planetary system.

\section{Analysis}\label{sec:analysis}
To optimally characterize the star-planet system, we leveraged the properties that \name{} shares with the Hyades to then perform a global analysis of the star-planet system. 
\subsection{Association of \name{} with the Hyades Cluster}\label{subsec:Hyades}
Galactic space velocity values were calculated using \texttt{PyAstronomy} tools, following the methodology outlined in \cite{Johnson1987} and utilizing data from \textit{Gaia} DR3. 
The spatial center of the Hyades cluster is $(-44.77, +0.40, -16.24)$\,pc \citep{Roser2019}. This can be compared to the Cartesian Galactic coordinates of \name{}, measured as $(-36.47, -17.86, -16.56)$\,pc. \name{} is approximately 20.1\,pc from the core of the Hyades cluster.
It has been shown that the Hyades exhibits a dispersed structure, including extensive tidal tails that span over one hundred parsecs in size \citep{Roser2019}. 
The structure of the cluster has been studied in detail using 3D kinematic modeling, confirming that \name{} is a high-confidence member of the Hyades, with a membership probability exceeding 99\% \citep{Oh2020}.
The location of \name{} within the tidal tail \citep{Roser2019} accounts for its distance relative to the core members of the Hyades \citep{Lodieu2019}.

\subsection{\name{} Photometric Rotation Period}\label{sec:Rotperiod}
To accurately determine the photometric rotation period of \name{}, we analyzed the processed \tess{} data, as described in \S \ref{sec:data}. The data were binned to 30 minutes to reduce scatter, better capturing longer-period stellar rotation signals. 
We also removed contaminated parts of the light curve, which included the initial 0.2 days of Sector 5 and the first 0.2 days after the downlink of Sector 32.

\begin{figure*}[tbh!]
    \centering    \includegraphics[width=0.95\textwidth]{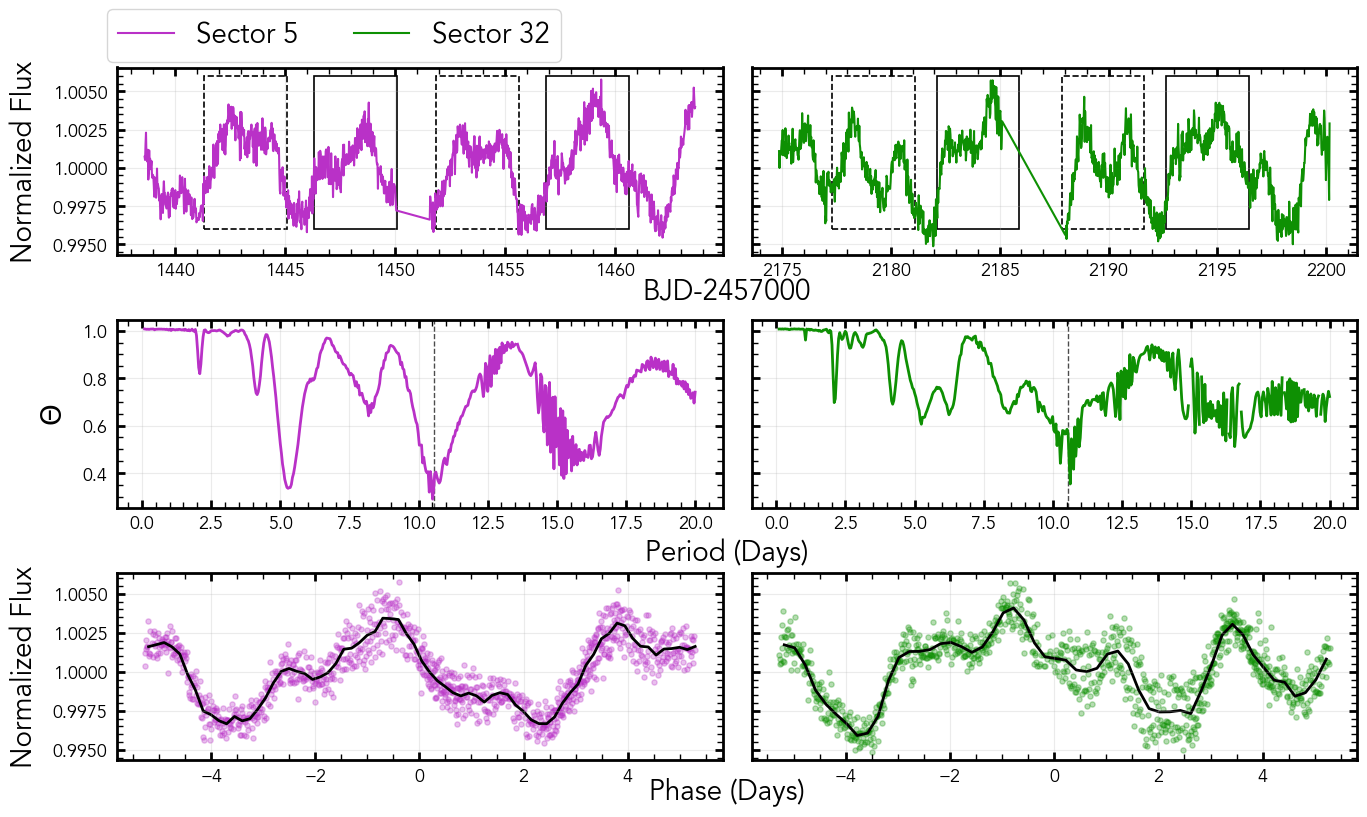}
    \caption{Analysis of \name{}'s photometric rotation period using \tess{} Sectors 5 (purple) and 32 (green).
    \textbf{Top:} Trimmed light curve with boxes showing similar structures; \textbf{Center:} PSM output, with maximum likely periods shown with dashed lines; \textbf{Bottom:} Phase-folded light curves for each sector at $P=$\protne{} days.}
   \label{fig:Rotresults}
\end{figure*}

Our Lomb-Scargle (LS) periodogram \citep{Lomb1976, Scargle1982} and autocorrelation function (ACF) analysis suggested a photometric rotation rate of approximately $5$\,days. However, under more scrutiny, we observed recurring substructure in the light curve indicating a ${\sim}10$\,day rotation period. This recurring substructure is illustrated in the top left and right panels of Figure~\ref{fig:Rotresults}. 

To optimally characterize this structure, we employed a phase dispersion minimization (PDM) algorithm from the package \texttt{starrotate}. 
The PDM algorithm detects periodic signals in unevenly sampled data by folding it over trial periods and minimizing the scatter within phase bins (described further in \cite{Stellingwerf1978}). The $\theta$ value reflects the scatter on the folded data, with values closer to zero having less scatter, with values closer to one having more scatter.    

The photometric rotation period is dependent upon the presence of spots, spot latitude, and the strength of differential surface rotation.
We measured a period of 10.51 days for Sector 5 and 10.63 days for Sector 32.
The PDM output and corresponding phase-folded light curves are shown in Figure~\ref{fig:Rotresults}. 
%This discrepancy in the performance of different methods of rotation analysis can be attributed to the spot location and frequency on the host star, which can often lead to harmonics of the actual period being mistakenly identified as the rotation period. As a young M dwarf, we would expect persistent spots to be a surface feature on \name{} \citep{Robertson2020}, explaining the problems encountered in the LS and ACF analysis. 

Stellar rotation offers a method to estimate a cluster's age. \cite{Skumanich1972} showed that average stellar rotational velocity decreases with cluster age. 
For low-mass stars, this decline is driven by magnetic braking, where the star’s magnetic field interacts with outflowing charged particles, removing angular momentum and slowing rotation over time \citep{Mestel1968, Gossage2023}.

\begin{figure}[tbh!]
    \centering    \includegraphics[width=0.48\textwidth]{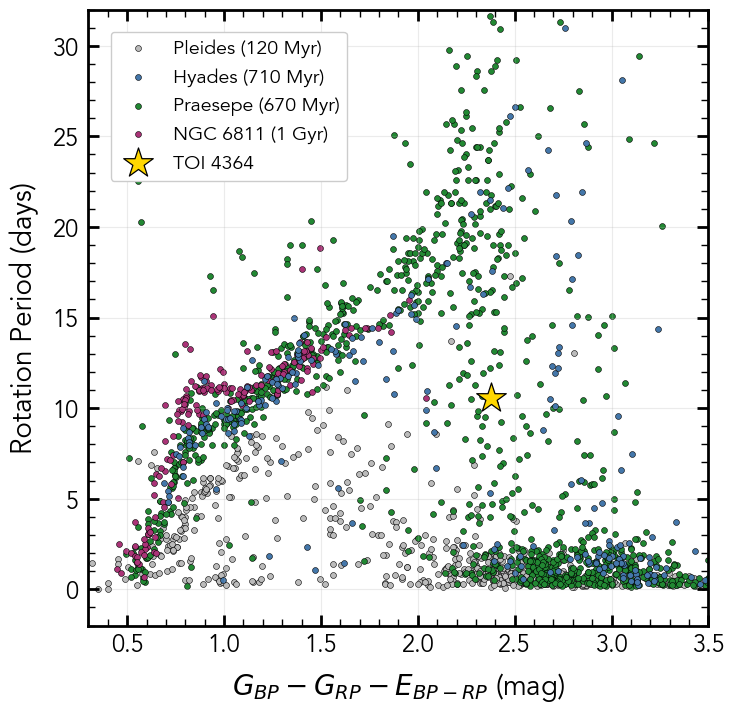}
    \caption{The color-rotation sequence for catalog members of the Hyades cluster, as compared to the Pleiades (120\,Myr; \citealt{Rebull2016}), Praesepe (670\,Myr; \citealt{Douglas2017}), and NGC 6811 (1\,Gyr; \citealt{Curtis2019}).
    \name{} is depicted as a gold star.}
   \label{fig:Hyadesrot}
\end{figure}

Consequently, the rotational sequence of a cluster can be compared with those of other clusters to estimate its age. Figure~\ref{fig:Hyadesrot} illustrates the rotation sequence for the Hyades alongside the Pleiades (120\,Myr), Praesepe (670\,Myr), and NGC 6811 (1\,Gyr). 
The rotational sequence of the Hyades exhibits a slope comparable to that of Praesepe \citep{Gossage2018}, thus supporting our age estimate of $710\pm 100$\,Myr.
Note that M dwarfs, like our planet host \name{}, exhibit a wide dispersion of rotation periods. 

\subsection{\texttt{EXOFASTv2} Analysis}
\label{sec:methods}

To determine the best-fit transit parameters, our team constructed a global fit from the \texttt{EXOFASTv2}\footnote{\url{https://github.com/jdeast/EXOFASTv2}} (\cite{Eastman2013, Eastman2019}) software package. 
\texttt{EXOFASTv2} utilizes a differential evolution Markov Chain Monte Carlo (MCMC) approach to fit both planetary and stellar parameters within a system simultaneously. 
This allowed us to jointly model the star's spectral energy distribution (SED) using \texttt{PARSEC} \citep{Bressan2012} in parallel with \tess{} and follow-up transit observations of the planet, thereby ensuring consistency across the data.

To construct the SED, we gathered archival photometric data from several sources: \Gaia{} \citep{GaiaMission, GaiaDR3}, the Two Micron All Sky Survey (2MASS; \cite{Cutri2003}), and the Wide-field Infrared Survey Explorer (WISE; \citealt{Wright2010, Cutri2012}). 
To set an upper bound for the visual extinction, we used the Bayestar 3D dust maps \citep{Green2015,dustmaps, Green2019} and we calculated a value of $A_{\mathrm{v}} = 0$. To account for this in the \texttt{EXOFASTv2} analysis, we then set an upper bound of $A_{\mathrm{V}} < 0.001$.   
In regard to previous estimates for the Hyades (see \cite{Brandner2023} for a summary), we opted to adopt a broad Gaussian prior for the age of \name{} of $700 \pm 100$\,Myr 
and a metallicity of [Fe/H]$=0.155\pm 0.095$\,dex.
%Drawing from the isochrone fits, we used a broad Gaussian prior for the age at $650 \pm 50$\,Myr, in line with estimates for the Hyades cluster, and a metallicity prior of $0.034$, as suggested by \cite{Perryman1998}. 
\begin{deluxetable}{lccc}[htb!]
\centering
\tabletypesize{\scriptsize}
\tablewidth{0pt}
\tablehead{\colhead{Parameter} & \colhead{Value} & \colhead{Source} }
\startdata
TIC ID & 4070275 & \tess{} Input Catalog\\
\textit{Gaia} DR3 ID &  3210444215030339584 & \Gaia{}\\
%& TOI-4364 & \cite{Cannon1993}\\
\hline
\multicolumn{3}{c}{Astrometry}\\
\hline
$\alpha$.  & $80.067\pm 	0.014 $ & \Gaia{} \\
$\delta$. & $-4.239\pm 0.011$ & \Gaia{} \\
$\mu_\alpha$ (mas\,yr$^{-1}$) & $64.456\pm 0.017$ & \Gaia{} \\
$\mu_\delta$ (mas\,yr$^{-1}$) & $43.789	\pm 0.014$ & \Gaia{}\\
$\pi$ (mas) & $22.805 \pm 0.012$ & \Gaia{}\\
distance (pc) & \dist{} & \Gaia{}\\
\hline
\multicolumn{3}{c}{Photometry}\\
\hline
Spectral Type & M4V & This paper \\
$\mathrm{G_{Gaia}}$ (mag) & \gaiamag{} & \Gaia{}\\
%describes why a Gmag error is not a single value https://gea.esac.esa.int/archive/documentation/GDR2/Gaia_archive/chap_datamodel/sec_dm_main_tables/ssec_dm_gaia_source.html
$\mathrm{G_{BP}}$ (mag) & $13.796 \pm 0.033$ & \Gaia{}\\
$\mathrm{G_{RP}}$ (mag) & $11.420 \pm 0.012$ & \Gaia{}\\
V (mag) & $13.56 \pm 0.03$ & \cite{Zacharias2013} \\
J (mag) & $9.952\pm 0.024$ & 2MASS\\
H (mag) & $9.307\pm 0.022$  & 2MASS\\	
Ks (mag) & $9.075\pm 0.024$ & 2MASS\\
W1 (mag) & $8.949\pm 0.023$ & ALLWISE\\
W2 (mag) & $8.857\pm 0.02$ & ALLWISE\\
W3 (mag) & $8.764\pm 0.03$ & ALLWISE\\ 
W4 (mag) & $8.833\pm 0.438$ & ALLWISE\\ 
$A_{\mathrm{V}}$ (mag) & $0.00050 \pm 0.00034$ &This paper$^*$ \\
\hline
\multicolumn{3}{c}{Kinematics \& Galactic Position}\\
\hline
RV$_{\rm{Bary}}$  (km\, s$^{-1}$) & $44.770 \pm 0.455$ & \Gaia{}\\
U (km\, s$^{-1}$) & $-43.505 \pm 0.047$ & This paper \\
V (km\, s$^{-1}$) & $-19.235 \pm 0.018$ & This paper \\
W (km\, s$^{-1}$) & $-2.002\pm 0.015$ & This paper \\
X (pc) & $-36.469\pm 0.039$ & This paper \\
Y (pc) & $-17.855\pm 0.016$ & This paper \\
Z (pc) & $-16.555\pm 0.003 $ & This paper \\
\hline
\multicolumn{3}{c}{Physical Properties}\\
\hline
$P_{\rm{rot}}$ (days) & $10.57 \pm 0.06$ & This paper\\
%\vsini \ (km\,s$^{-1}$) & $12.16 \pm \tbd{XX}$ & This paper \\
%$i_*$ ($^\circ$) & $\approx 90$ & This paper \\
Age (Myr) & \age{} & This paper\\
T$_{\mathrm{eff}}$ (K) & $3528^{+33}_{-20}$ & This paper \\
$\mathrm{M}_\star$ (\Msun{}) & \mass{} & This paper$^*$\\
R$_\star$ (\Rsun{}) & \rad{} &This paper$^*$ \\ 
L$_\star$ (\Lsun{}) & $0.03075^{+0.00085}_{-0.00084}$ & This paper$^*$\\
L$_X$ (erg \,$s^{-1}$) & $2.07 \times 10^{28}$ & \cite{Freund2020} \\
$\rho_\star$ ($\mathrm{g} \, \mathrm{cm}^{-3}$) & $6.69^{+0.38}_{-0.33}$ & This paper$^*$\\
$\log{(g)}$ $(\log(\mathrm{cm/s^2})$ & $4.785^{+0.017}_{-0.016}$ & This paper$^*$ \\
$\mathrm{[Fe/H]}$ (dex) & \feh{} &This paper$^*$ \\
\enddata
\caption{Stellar properties of \name{}. Values marked with an asterisk were obtained from \texttt{EXOFASTTv2}.\label{tab:prop}}
\end{deluxetable}

The planetary parameters are predominately constrained by the transit light curves listed in Tables~\ref{tab:tessphot}-\ref{tab:fuobs}. The \tess{} light curves were processed using the method discussed in \S \ref{subsec:tess}. They were then flattened using \texttt{keplersplinev2},\footnote{\url{https://github.com/avanderburg/keplersplinev2}} a Python-based spline fitting tool. The spline breakpoint spacing was chosen as the value that minimizes the Bayesian Information Criterion (BIC). 
The \tess{} light curves were then truncated to remove unnecessary out-of-transit data, retaining a baseline of one transit duration on either side of each transit. 
The follow-up light curves were flattened using \texttt{AstroImageJ} \citep{Collins2017}.
No prior estimate was placed on the linear and quadratic limb darkening coefficients, which were allowed to vary as free parameters. To estimate the mass of the planet, \texttt{EXOFASTv2} uses the mass-radius relation from \cite{Chen2017}.

We followed the default \texttt{EXOFASTv2} MCMC setup, with twice as many chains as the number of parameters in each fit. For the circular fit, this meant 46 parameters and 92 chains. For the fit where eccentricity was allowed to float, there were 50 parameters and 100 chains. 
To ensure robust and converged results, we implemented strict convergence criteria, requiring a Gelman-Rubin statistic \citep{Gelman1992} less than 1.01 and more than 1000 independent draws. The Gelman-Rubin statistic criterion is typically the more difficult to reach. In the case of the eccentric fit, the MCMC had a Gelman-Rubin statistic of 1.0084 and 5783 independent draws after accepting 9.26\% of the total number of steps. 
Our \texttt{EXOFASTv2} MCMC circular fit resulted in a Gelman-Rubin statistic of 1.0041 and 4727 independent draws after accepting 17.02\% of the total number of steps.
The resulting stellar parameters are provided in Table~\ref{tab:prop}. 
The orbital fit parameters are provided in Table~\ref{tab:fitparams}.
The best-fit models are illustrated in Figures~\ref{fig:MCMCtransitfit}-\ref{fig:FUtransitfit}.

\begin{deluxetable}{ccc}
\large
\centering
\tablewidth{0pt}
\tablehead{
\colhead{Parameter} & \colhead{Value $e$ free}& \colhead{Value $e=0$ }
}
\setlength{\tabcolsep}{15pt}
\startdata
$P$ (days)   &$5.424019^{+0.000010}_{-0.000011}$& $5.424019\pm0.000011$\\
$T_0$ (BJD)  &$2458439.3428 \pm 0.0020$ & $2458439.3431\pm0.0020$\\
$a/R_{\star}$ & $21.83^{+0.41}_{-0.36}$ & $21.86^{+0.41}_{-0.36}$\\
$i$ (degrees) &$88.68^{+0.85}_{-0.40}$& $88.46^{+0.11}_{-0.10}$ \\
$e$  & $0.29^{+0.38}_{-0.17}$ & 0\\
$\omega_{\star}$ (degrees) &$-190^{+160}_{-130}$ & 0\\
$u_{1}$ (\textit{TESS}) & $0.312\pm0.036$& $0.287\pm0.036$\\ 
$u_{2}$ (\textit{TESS}) & $0.34\pm0.036$& $0.360^{+0.036}_{-0.035}$\\   %&$0.360\pm0.035
$u_{1}$ ($i'$) & $0.341 \pm 0.021$ & $0.316^{+0.019}_{-0.020}$\\
$u_{2}$ ($i'$) & $0.315^{+0.020}_{-0.021}$ &$0.336\pm0.019$\\   
$R_{\mathrm{p}} /R_{\star}$ &$0.0394^{+0.0017}_{-0.0014}$ & $0.0402^{+0.0012}_{-0.0013}$ \\
%$\delta = \left(R_{\mathrm{p}} /R_{\star}\right)^2$ &$0.00155^{+0.00014}_{-0.00011}$ \\
%~~~~$\delta_{\rm i'}$\dotfill &Transit depth in i' (fraction)\dotfill &$0.00177\pm0.00011$\\
$\delta_{\rm \textit{TESS}}$ & $0.00178\pm0.00010$ & $0.00175\pm0.00010$\\
$\delta_{\rm i'}$ & $0.00179 \pm 0.00010$ & $0.00180\pm0.00010$\\
$t_{14}$ (days) &$0.0668^{+0.0020}_{-0.0015}$& $0.0677^{+0.0014}_{-0.0013}$\\
$t_{12}$ (days) & $0.00298^{+0.0013}_{-0.00047}$ & $0.00393^{+0.00023}_{-0.00022}$\\ 
%$logg_P$\dotfill &Surface gravity$^{4}$ \dotfill &$3.11^{+0.13}_{-0.11}$\\
%~~~~$\fave$\dotfill &Incident Flux (\fluxcgs)\dotfill &$0.0174^{+0.0015}_{-0.0046}$\\
%~~~~$M_{\mathrm{p}} \sin i$\dotfill &Minimum mass$^{4}$ (\mj)\dotfill &$0.0170^{+0.0063}_{-0.0040}$\\
%~~~~$M_{\mathrm{p}} /M_*$\dotfill &Mass ratio$^{4}$ \dotfill &$0.0000332^{+0.000012}_{-0.0000077}$\\
%~~~~$d/R_{\star}$\dotfill &Separation at mid-transit \dotfill &$17.6^{+4.0}_{-3.6}$\\
\hline
\multicolumn{3}{c}{Derived Parameters}\\
\hline 
$R_{\mathrm{p}}$ (\RE{})  &\prad{}  & $2.05^{+0.08}_{-0.08}$ \\
$a$ (AU) &$0.04753^{+0.00053}_{-0.00054}$& $0.04753\pm0.00053$\\
$T_{eq}$ (K) &\teq{} & $534.4 \pm 3.9$\\
%$M_{\mathrm{p}}$ (\ME)  &\pmass  &  $5.53^{+1.9}_{-1.3}$ \\
%$\rho_P$ (g\,cm$^{-3}$)&$3.58^{+1.3}_{-0.82}$ & $3.52^{+1.3}_{-0.80}$ \\
%$\tau_{\rm circ}$ (Gyr)  &$92^{+120}_{-91}$ & - \\
%$K$ (m\,s$^{-1})$& $3.55^{+1.8}_{-0.93}$ & $3.25^{+1.1}_{-0.77}$\\
TSM &  \tsm  & 45.1  \\
ESM & \esm  & 5.1 \\
\enddata
\label{tab:fitparams}
\caption{Transit-Fit Parameters for \pname{}. \textbf{Top:} Transit fits obtained from the joint SED, \tess{}, and follow-up photometry using the \texttt{EXOFASTv2} model. \textbf{Bottom:} Derived system parameters. The middle column presents the results when the eccentricity ($e$) and argument of periapsis ($\omega$) are allowed to vary as free parameters. The rightmost column shows the fit assuming $e = 0$ and $\omega = 0$.}
\end{deluxetable}

\section{Ruling out False Positive Scenarios}
\label{sec:Vetting}
\subsection{Possible False Positive Scenarios}
In our efforts to validate the detection of \pname{}, we utilized \texttt{LEO-vetter},\footnote{\url{https://ascl.net/2404.026}} an automated tool designed to assess potential exoplanet signals in light curve data. \pname{} successfully passed all assessments conducted by \texttt{LEO-vetter}, including flux and pixel vetting analyses. These tests examined the transit shape to exclude eclipsing binaries and analyzed centroid offsets to confirm the accuracy of the transit event locations. 

Given the absence of additional validation methods for \pname{}, we worked to exclude all plausible false positive scenarios, as detailed below.

\begin{itemize} \item \textbf{Instrumental Artifacts and Stellar Variability:} 
No instrumental artifacts in the \name{} \tess{} light curves have been identified that could replicate the observed signal in the BLS periodogram. Furthermore, the transit signal has been independently verified photometrically by observations from four separate facilities, eliminating the likelihood of an instrumental artifact. While M-dwarfs are known for their significant variability \citep{Mignon2023}, particularly those younger than 1\,Gyr, no known phenomenon could induce a transit-like flux dip that recurs consistently over multiple years. Stellar variability typically manifests as gradual changes over days, weeks, or months, rather than the distinct transit-like flux dips observed.

\item \textbf{Target is an Eclipsing Binary}: 
A possible concern is that the target star might be an eclipsing binary (EB). However, if this were the case, one would expect large-scale RV variations on the order of kilometers per second, as the companion would need to have an inclination of near $90$ degrees to produce a transit-like signal. Instead, the \textit{TRES} data reveals a maximum spread between RVs to be 78\,$\mathrm{m \, s^{-1}}$ (taken at opposite quadratures), indicating that \name{} is not an EB in a 5.42 day orbit. 

\item \textbf{Contamination by Foreground or Background Eclipsing Binaries}: 
\begin{figure}%[tbh!]
   \centering
   \includegraphics[width=0.48\textwidth]{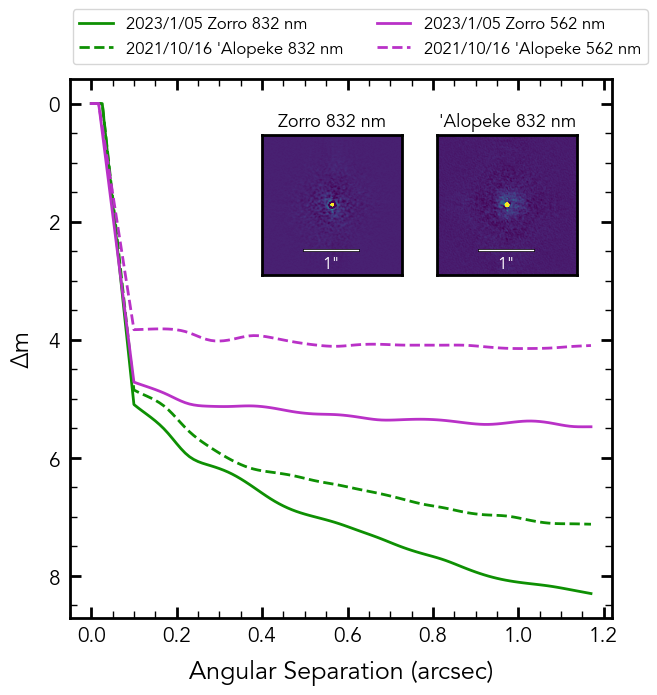}
   \includegraphics[width=0.47\textwidth]{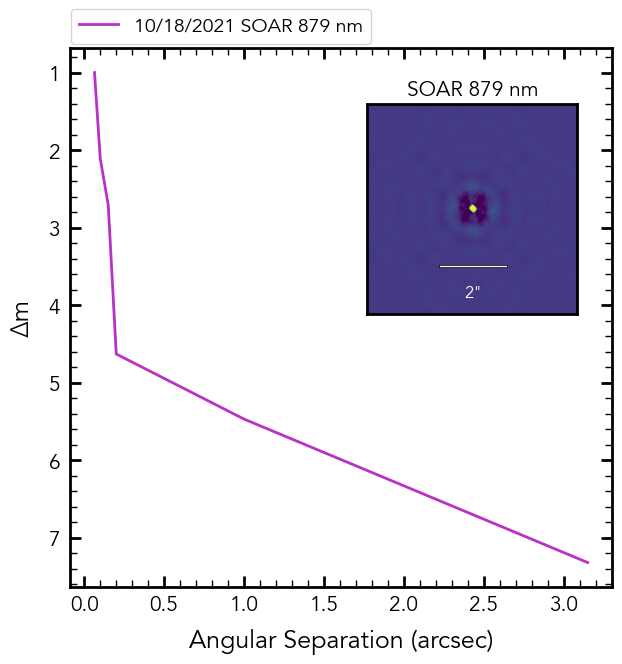}
    \caption{\textbf{Top Panel:} Contrast curves obtained from Gemini North ‘Alopeke and Gemini South Zorro instruments. Insets showcase 832\,nm reconstructed images, highlighting the increased contrasts achieved with Zorro on Gemini South due to better seeing conditions and a longer total integration time. No close companion star was detected within the achieved 5$\sigma$ magnitude contrast curves and angular limits from the 8-m diffraction limit out to 1.2\,\arcsec. At the distance of TOI-4364, these limits correspond to spatial separations ranging from $0.88-52.8$\,AU. 
    \textbf{Bottom Panel:} SOAR sensitivity curve taken on October 18, 2021. The $5\sigma$ sensitivity limits are shown out to 3.15\arcsec{}, corresponding to a projected separation of 138.1\,AU.}
  \label{fig:contrast}
  % 879 (289)
\end{figure}
Another potential false positive scenario is the presence of an EB in the foreground or background of the target, which would not have an effect on the RV measurements of \name{}.
Given the large pixel scale of \tess{}, this is an important consideration.  

The required magnitude difference of a contaminating EB is given by
\begin{equation} \Delta m \lesssim 2.5 \log_{10} \left(\frac{t_{12}^2}{t_{13}^2 \delta}\right), \end{equation} where 
$\delta$ is the transit depth, $t_{12}$ is the ingress duration,  $t_{14}$ is the full transit duration, and $t_{13} = t_{14} - t_{12}$ \citep{Vanderburg2019}. 

% Instead of using the \textit{EXOFASTv2} parameters, which effectively predicts the ingress/egress time based on the stellar density, 
To calculate this magnitude difference, our team employed the differential MCMC code \texttt{edmcmc}\footnote{\url{https://github.com/avanderburg/edmcmc}} coupled with \texttt{batman}\footnote{\url{https://lkreidberg.github.io/batman/docs/html/index.html}} to optimally constrain the orbital parameters. We employed the data, as described in \S\ref{subsec:tess}, removing observations that were not within $\pm0.1$\,days of the transit (as predicted by \texttt{EXOFASTv2}). We ran the code with 50 walkers for 10,000 links, with 4,000 burn-in links. We fixed the eccentricity to zero and argument of periapsis to $90^{\circ}$ and used a prior of the \texttt{EXOFASTv2} output for the walkers. All orbital parameters were run to a Gelmin-Rubin statistic of $<1.05$. Following this analysis, we arrive at a limiting contaminating EB magnitude of $\Delta m \le 3.65$\,mag at 95\% confidence.  
heZor 832\,nmro speckle imaging contrast curves in Figuthe top panel of re~\ref{fig:contrast} illustrate a 3.87g contrast at 0.085", corrwhich esponding sa projected separation of 3.73\,AU. 
For the outer limit, the 879\,nm \textit{SOAR} imaging observation (bottom panel of Figure~\ref{fig:contrast}) also confirmed the lack of an object bright enough to cause this signal out to 3". 
Although this is within the follow-up photometric radius, a 15.04 \tess{} magnitude star would be clearly resolved by \textit{Gaia} given a contrast sensitivity above $99\%$ for $\Delta G_{\mathrm{Gaia}} < 6$\,mag \citep{Brandeker2019}. Given that the follow-up photometry constrained the signal to be within 4.0" of \name{}, it is very unlikely the signal comes from a contaminating EB.

\item \textbf{Physically Associated Eclipsing Binaries}: 
Another possibility is that \name{} has a physically associated EB companion that could account for the observed signal. This cannot be completely ruled out, as the angular separation may be small enough such that it is unresolved by the speckle imaging.
In such a case, the inclination would need to be near-zero degrees to prevent a significant RV shift in \name{}. This is unlikely for several reasons. Firstly, it is statistically unlikely given that M dwarfs have a very low triple rate of $\approx3.3 \%$  \citep{Winters2019}).
Secondly, the Zorro contrast curve rules out the presence of companions bright enough to induce this signal at distances beyond 3.73\,AU (corresponding to a maximum period of 10.3 years).
Given the maximum of \textit{ESPRESSO} RV shift of 28\,m s$^{-1}$ shift (taken over three months), a companion would need to have an extremely low inclination. 
While there is no way to completely rule out this scenario, it is unlikely that a bound EB is producing the transit signal of \pname{}.
\end{itemize}

\subsection{Statistical Validation with \texttt{TRICERATOPS}}
\label{sec:triceratops analysis}
To further validate our conclusion that no false positive scenarios could mimic the signal of \pname{}, we employed the \texttt{TRICERATOPS} pipeline \citep{Giacalone2021} to assess the likelihood of various false positive scenarios. \texttt{TRICERATOPS} utilizes Bayesian inference to compute the probability that an input light curve originates from a wide range of scenarios.
This includes assessing the shape of the transit signature and conducting checks for instrumental artifacts or background contamination. 

We ran \texttt{TRICERATOPS} twenty times using data from the contrast curves obtained by the Zorro speckle imaging. Each run included $10^6$ instances, which yielded a false positive probability --- the likelihood that the transit signal is not due to a planet --- of $0.00319 \pm 0.00017$. 
Similarly, the nearby false positive probability --- the likelihood that the transit signal is due to a nearby object --- was found to be $0.00317 \pm 0.00017$. These low false alarm probabilities allow us to statistically confirm that \pname{} is indeed a planet.
Moreover, while the false alarm probabilities are low, they likely overestimate the true values since existing follow-up photometry using smaller apertures could not be included. From this analysis, we conclude that \pname{} is a genuine exoplanet.

\section{Discussion}\label{sec:discussion}
\subsection{\pname{} in Context}
\pname{} serves as a key benchmark for studying exoplanets within the Hyades cluster, which is known to host eight planets. Table~\ref{tab:Hyads} presents a comparison of \pname{}'s properties with those of the other Hyades planets. Notably, \pname{} has the second shortest orbital period among them. 
Short-period planets like in clusters are important follow-up targets, amenable to further constraints and characterization.
Further, its location near the upper boundary of the radius valley makes it a particularly valuable target for probing planetary evolution. Given its youth and proximity, \pname{} offers an invaluable opportunity to investigate how planetary radii evolve as a function of system age.

\begin{deluxetable}{lrrrr}[htb!]
\centering
%\LARGE
\tabletypesize{\scriptsize}
\tablewidth{0pt}
\caption{Planets in the Hyades
\label{tab:Hyads}}
\tablehead{\colhead{Planet} & \colhead{Period (d)} & \colhead{Radius (\RE)} & \colhead{Distance (pc)} }
\startdata
\textbf{\pname{}} & \textbf{5.4} & \textbf{2.0} & \textbf{43.9}  \\
K2-25\,b & 3.5  & 3.4 &  45.0  \\
HD 285507\,b & 6.1 &13.8 & 45.0  \\
K2-136\,b & 8.0  & 1.0& 59.2  \\ 
K2-136\,c & 17.3 & 2.9  & 59.2 \\ 
K2-136\,d & 25.6 & 1.5 &  59.2 \\
HD 283869\,b & 106.0 & 2.0 & 47.5 \\
HD 28305\,b & 594.9 & -- & 44.7 \\ 
\enddata
\end{deluxetable}

\begin{figure}[htb!]
    %\centering
  \includegraphics[width=0.49\textwidth]{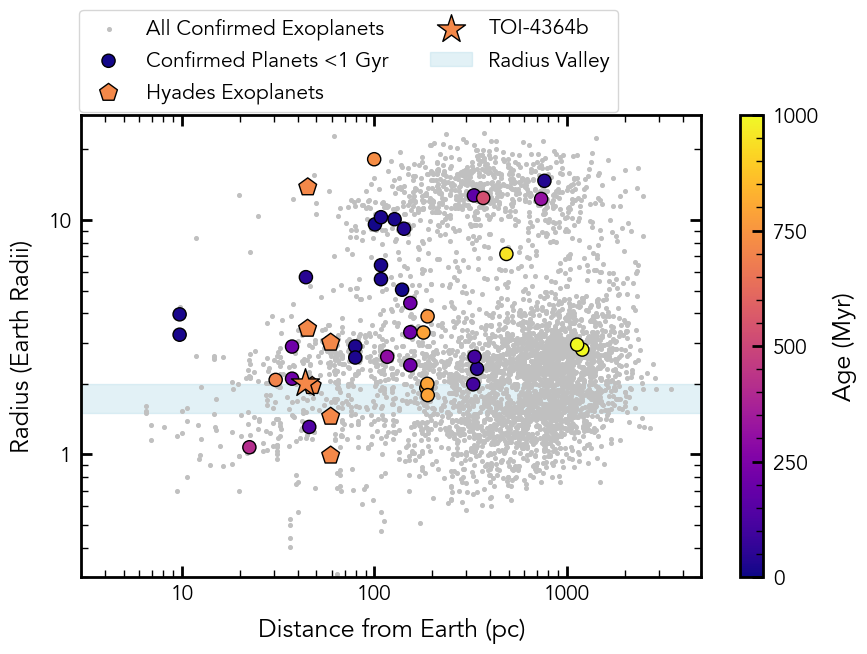}
    \caption{Distance from Earth versus planet radius for confirmed exoplanets (grey circles), young ($<1$\,Gyr) planets with well-constrained ages (colored circles), previously discovered Hyades exoplanets (pentagons), and \pname{} (star). \pname{} stands out as a nearby young planet at the upper edge of the radius valley, making it a prime target for atmospheric and evolutionary studies.}
   \label{fig:distanceversusradius}
\end{figure}

When compared to the broader population of confirmed exoplanets, as shown in Figure~\ref{fig:distanceversusradius}, \pname{} continues to stand out for its proximity to Earth. This is especially significant when considering planets younger than $1$\,Gyr with well-constrained ages.
For clarity, we define a ``well-constrained" age as one where both the upper and lower error is at most 20\% of the estimated value.
The apparent brightness of \pname{} make it an exceptional target for studying planetary evolution among M dwarf planetary companions. Precise age constraints for M dwarf planets are often lacking due to the host stars' low luminosity and the challenges associated with dating these slowly-evolving stars. 
The proximity of \pname{}, combined with its well-constrained age, provides a rare opportunity to address this gap, offering critical insights into how planets form, migrate, and evolve around low-mass stars.

\begin{figure}[tbh!]
    \centering    \includegraphics[width=0.49\textwidth]{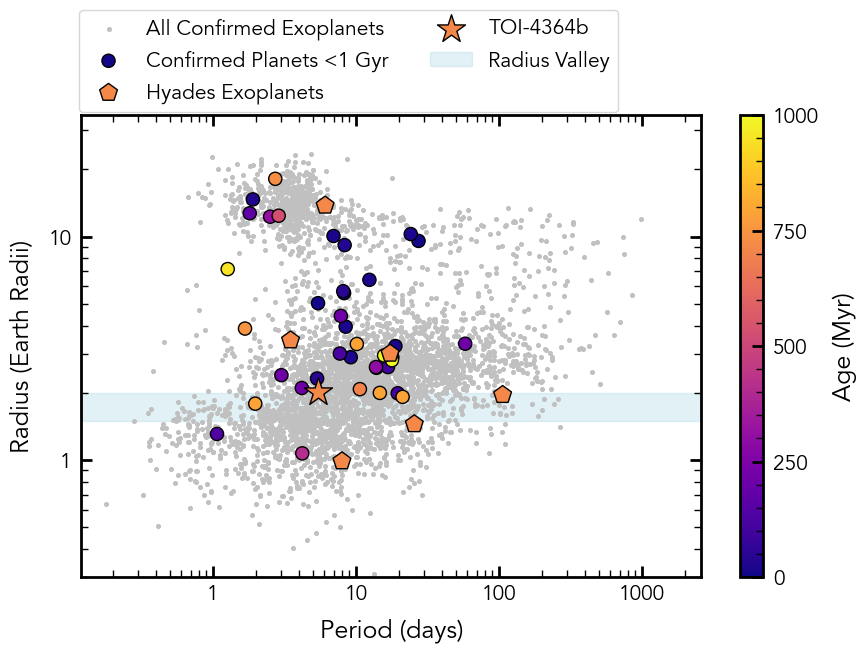}
    \caption{Period-radius distribution of confirmed exoplanets (grey circles),  confirmed young ($<1$\,Gyr)  exoplanets with well-constrained ages (colored circles), previously discovered Hyades exoplanets (pentagons), and \pname{} (star).
    To enhance the readability of the figure, we omit planets discovered via direct imaging. Note that \pname{} resides at the upper edge of the radius valley (blue strip).}
   \label{fig:periodvradius}
\end{figure}

Figure~\ref{fig:periodvradius} illustrates the period-radius diagram, highlighting young objects with precisely measured age estimates, including \pname{} and other known Hyades exoplanets.
Even among young systems, \pname{} currently resides at a critical boundary of the radius valley.

\subsection{Evolving Physical and Orbital Properties}
Comparisons to older planets and broader planet samples must be contextualized with the ongoing evolutionary changes that are relevant to a young planet like \pname{}.
In the following subsections, we discuss some of these considerations.

\subsubsection{Planetary Radius}
The measured radius of \pname{} of \prad{}  places it just above the threshold where a planet is expected to possess an H/He envelope \citep{Lopez2014, rogers}. 
Yet, given the young age of this planet, it is important to note that the radius is likely to change with time, due to both cooling and photoevaporation from the insolation by the host star. Younger planets will retain higher levels of envelope entropy, and as planets with H/He envelopes are irradiated by their host stars, they will also lose mass due to photoevaporation. 
This mass loss rate \citep{Watson1981,Lopez2012, Luger2015, Barnes2020} can be modeled as:
\begin{equation}
\dot{m} = \epsilon \left(\frac{{R_p}^{3} L_{\mathrm{XUV}}}{4 K  a^2 G M_{p}}\right),
\label{eq:mdot}
\end{equation}
where $\epsilon =0.1$ is the heating efficiency, $R_{\mathrm{p}}$ is the radius of the planet, $L_{\mathrm{XUV}}$ is the XUV luminosity of the host star, $K$ is the tidal enhancement factor \citep{Erkaev2007}, $a$ is the semi-major axis, $G$ is the gravitational constant, and $M_p$ is the mass of the planet (predicted from \citealt{Chen2017} to be \pmass{}). 
We assume that the planetary atmosphere becomes optically thick to XUV photons at 1\,$R_{\mathrm{p}}$. In the subsequent analysis, we set all planetary parameters to be their best-fit values given in Table~\ref{tab:fitparams} for the free-eccentricity solution, and assume a stellar mass of $0.49$\,\Msun{}.
%\khalid{Since we have mass measurement, a mass-radius diagram is needed to show the position of the planet and theoretical model from Lopez and Rogers.}

To evaluate the radius evolution of \pname{},
we ran a series of models using the code \vplanet\ \citep{Barnes2020} and its \atmesc\ and \stellar\ modules. \vplanet\  allows us to integrate the effects of photoevaporation and planetary cooling to simulate the evolution of a planet's radius as a function of stellar age, planet mass, and envelope fraction. In our models, we use interpolated stellar luminosity tracks from \citet{Baraffe2015} and the mass-radius model from \citet{Lopez2012}.
Our simulations assume that photoevaporation begins after disk dissipation at approximately 10\,Myr.

\begin{figure}[htb!]
    \centering
    \includegraphics[width=0.45\textwidth]{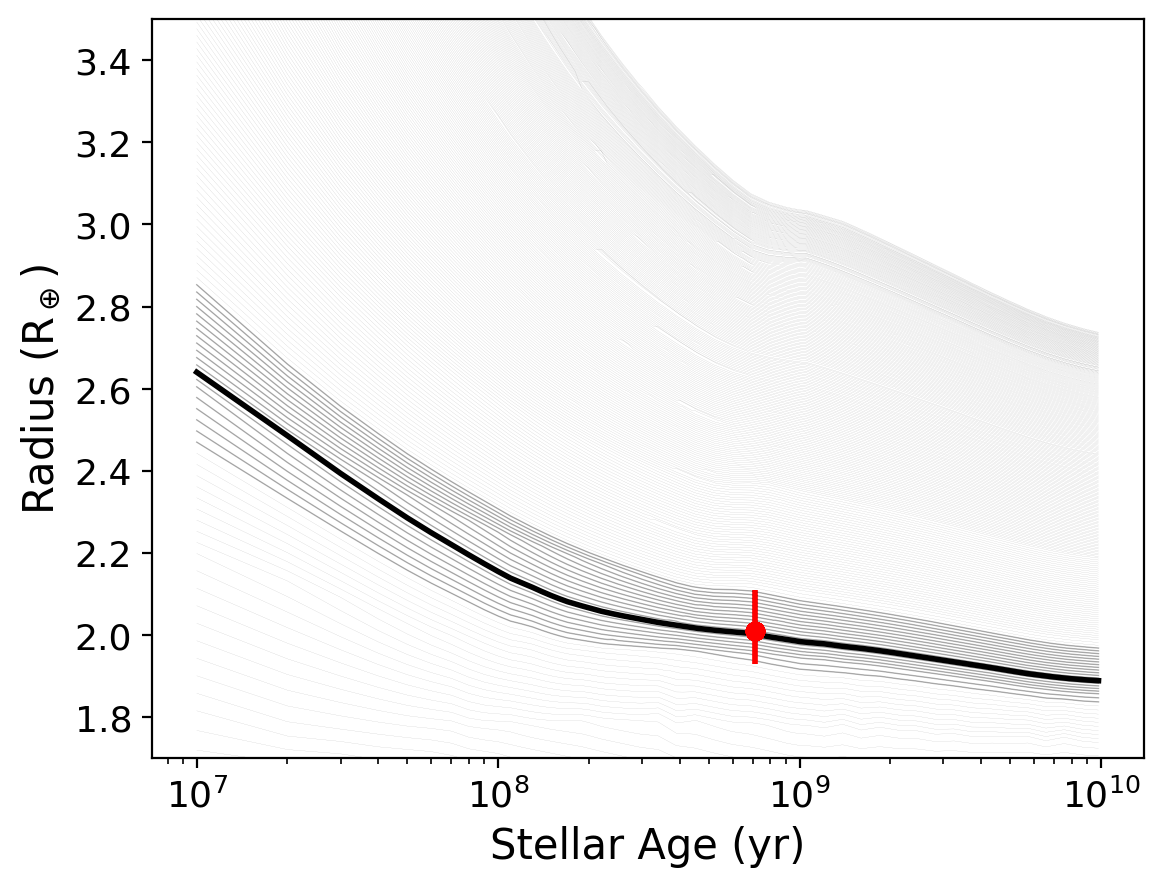}
    \caption{Evolution of the planetary radius as a function of stellar age for \pname{}, assuming the best-fit planetary parameters (including planetary mass). The grey and black lines represent the radius evolution predicted by the \vplanet\  model, which incorporates both planetary cooling and photoevaporation effects. Each line corresponds to a different initial H/He envelope fraction, ranging from $0\%-10$\%. The red point marks the observed planetary radius at 710\,Myr, with error bars indicating the 1$\sigma$ uncertainty. The darker contour lines show models that are consistent with the observed radius, which have H/He envelope fractions between $0.4\%-0.9\%$. The nominal best-fit solution is shown as a thicker line, which corresponds to a current-day envelope fraction of 0.6\%. Over time, the planet's radius is expected to decrease gradually due to continued cooling and mass loss from photoevaporation.}
   \label{fig:radiusevolution}
\end{figure}

In the first set of simulations, we set the planet mass to the predicted mass from \citealt{Chen2017}, running 300 simulations with initial H/He envelope mass fraction varying between $0\%-10\%$. 
The results of these simulations are shown in Figure \ref{fig:radiusevolution}, which depicts the computed evolution for each of our 300 grid points. 
For comparison to the models, the current-day measured radius of \pname{} is denoted by a red point with error bars denoting the 1{$\sigma$} uncertainty. 
Models consistent with the observed radius of \pname{} are denoted by darker, thicker lines, with the nominal best-fit solution shown by the largest line width. 
Models consistent with the observed radius at 710\,Myr, correspond to current H/He envelope fraction ranging between $0.4\%-0.9\%$.
This corresponds to an initial H/He envelope mass fraction range of $0.8\%-1.4\%$. 
The best-fit solution suggests an initial envelope fraction of 1.0\%, which, after accounting for the aforementioned evolutionary processes, results in a current H/He envelope fraction of 0.6\%. Over the next few Gyr, the predicted radius evolution for these allowed solutions is minimal, with a maximal 10\% decrease.

We ran a second set of simulations to assess the impact of the uncertainty of the planetary mass on the computed allowable envelope fractions. In the second set of simulations, we varied both the planet’s mass (within the 1$\sigma$ range) and its initial envelope fraction, with a total of 20 grid points in each dimension for a total of 400 simulations. 
This resulted in a broader range of viable initial H/He envelope mass fractions.
These span between $0.5\%-1.8\%$, with current envelope mass fractions ranging between $0.4\%-1\%$. 
In this case, the predicted radius evolution is more substantial, with the planet’s radius potentially decreasing by up to 15\% over the next few Gyr. We note that these envelope fractions assume the planet's true mass lies within the 1$\sigma$ range given by \citealt{Chen2017}.

\subsubsection{Tidal Locking}
Understanding whether \pname{} is tidally locked to its host star is crucial for accurate interpretation of the planet's atmospheric circulation patterns, atmospheric evolution (including the presence of enhanced atmospheric stripping), and thermal evolution. 
More specifically, when tidal forces synchronize a planet's rotation to the frequency of its orbit, the day side of the planet receives \textit{all} of the incident radiation, which directly impacts the aforementioned atmospheric properties \citep[e.g.,][]{Pierrehumbert2019, Griebmeier2004}.  

For short-period planets, tidal forces will tend to lock the planetary spin frequency to the planetary mean motion on a timescale given by
\begin{equation}
    t_{lock} \approx \frac{\omega a^6  I Q}{3 G M_{\star} ^2 k_2 R_{\mathrm{p}}^5},
\end{equation}
where $\omega$ is the planet's initial rotation rate, $I$ is the moment of inertia of the planet, $Q$ is the planet's tidal dissipation factor (a dimensionless number that describes how efficiently tidal energy is dissipated within the planet), and $M_{\star}$ is the mass of the host star, $k_2$ planet's Love number (which characterizes how much the planet deforms in response to tidal forces) \citep{Gladman1996}.

Instead of solving for the tidal locking time by assuming $Q$, one can provide an age constraint and determine the minimum $Q$ required for the system to not be locked. 
Assuming an initial rotation rate of one cycle per Earth day, a $Q \gtrsim 10^8$ is required to tidally lock the planet within 710\,Myr. 
All the Solar System planets have values much lower than this, including Uranus and Neptune ($Q \approx 10^5$, \citealt{Tittemore1990}).
Similarly, terrestrial planets have values much lower than this ($Q \approx 10 - 300$, \citealt{Goldreich1966}). 
As such, we can reasonably conclude that this planet is tidally locked.

\subsection{Metrics for Atmospheric Characterization}
\label{subsec:metrics}
Characterizing a planet's atmosphere is crucial for understanding its composition and structure.
Two key metrics used to assess exoplanet atmospheres --- particularly in preparation for observations with next-generation telescopes like the James Webb Space Telescope (JWST) --- are the Transmission Spectroscopy Metric (TSM) and the Emission Spectroscopy Metric (ESM). 
These metrics help prioritize targets for detailed follow-up studies based on the planet’s ability to reveal atmospheric properties through transit and emission observations.

A fundamental parameter influencing both the TSM and ESM is the planet’s equilibrium temperature, which provides a baseline for estimating atmospheric thermal structure and potential for atmospheric escape. For \pname{}, this temperature gives valuable insight into the energy balance of the planet and its atmospheric stability. By calculating these metrics, we can assess \pname{}'s value as a target for transmission and emission spectroscopic follow-up. Assuming an albedo ($\alpha$) of 0.3, and perfect heat redistribution, the equilibrium temperature is approximated by
\begin{equation}
    T_{\mathrm{eq}} = \Bigg(\frac{L_{\star} (1-\alpha)}{16 \pi \sigma a^2 }\Bigg)^\frac{1}{4},
\end{equation}
where $L_{\star}$ is the host star's luminosity and $\sigma$ is the Stefan-Boltzmann constant.
We calculate an equilibrium temperature of \teq{}\,K for \pname{}.

The TSM evaluates the planet's suitability for transmission spectroscopy, where the atmosphere is observed as the planet passes in front of its host star, as defined in \cite{Kempton2018}. 
For \pname{}, we derive a TSM of \tsm{}, indicating that this is a modest candidate for transmission observations when compared to the most promising TSM targets.
Yet, this value is particularly compelling when compared to the values of other young systems with well-constrained ages, as described in \S \ref{subsec:tsm}.

The ESM, which is also defined in \cite{Kempton2018}, measures the planet’s suitability for emission spectroscopy --- the detection of thermal radiation from the planet’s dayside when illuminated by its star. Using the formula from \cite{Cowan2011} for a tidally locked planet, the dayside temperature is given by
\begin{equation}
    T_{\mathrm{day}} = \bigg[\frac{2}{3} (1-\alpha)\bigg]^{\frac{1}{4}} T_{\mathrm{sub}}, 
\end{equation}
where $T_{\mathrm{sub}}$ represents the temperature at the substellar point (the point on the planet's surface such that the host star is at the zenith) and is given by
\begin{equation}
    T_{\mathrm{sub}} = T_{\mathrm{eff}} \bigg(\frac{R_{\star}}{a}\bigg)^\frac{1}{2},
\end{equation}
where $R_{\star}$ is the stellar radius. 
For \pname{}, we derive an ESM of \esm{}, also indicating this is a viable candidate for emission observations.
Once again, this is a particularly compelling target when compared to values of other young, well-constrained systems, as described in \S \ref{subsec:tsm}.
%\begin{equation}
%    ESM = 4.29\times 10^6 \times \frac{B_{7.5}(T_{\mathrm{day}})}{B_{7.5}(T_\star)}\times \bigg(\frac{R_{\mathrm{p}} }{R_\star} \bigg)^2 \times 10^{-(m_K/5)},
%\end{equation}
%where $B_{7.5}(T)$ is the Planck blackbody function evaluated at 7.5\,$\mu m$ (for the dayside temperature and the effective temperature of the star) and $m_K$ is the apparent magnitude of the star in the K band. 

\subsection{Future Work Characterizing \pname{}}
\subsubsection{Radial Velocity Observations}
To accurately determine the mass of \pname{}, high-precision RV measurements are essential. 
Assuming a planetary mass of \pmass{} from \cite{Chen2017}, we would expect an RV semi-amplitude of $3.55_{-0.93}^{+1.3}\, \mathrm{m}\,\mathrm{s}^{-1}$. 
Contemporary high-precision spectrographs, such as WIYN/NEID, ESPRESSO, the Habitable Zone Planet Finder, and Gemini/MaroonX have the technological capability of detecting RV variations of this magnitude. However, a young M dwarf is expected to have high amplitude stellar variability, potentially posing challenges for obtaining a mass measurement.

\subsubsection{Atmospheric Characterization}\label{subsec:tsm}

While the equilibrium temperature of \pname{} is too high to support life, \pname{} is a viable candidate for atmospheric characterization through transmission and emission spectroscopy. 
With a transit depth of 1.8\,ppt, a short orbital period, and the system's close proximity to Earth, this planet offers a unique opportunity for detailed follow-up observations.

\begin{figure}[tbh!]
\includegraphics[width=0.48\textwidth]{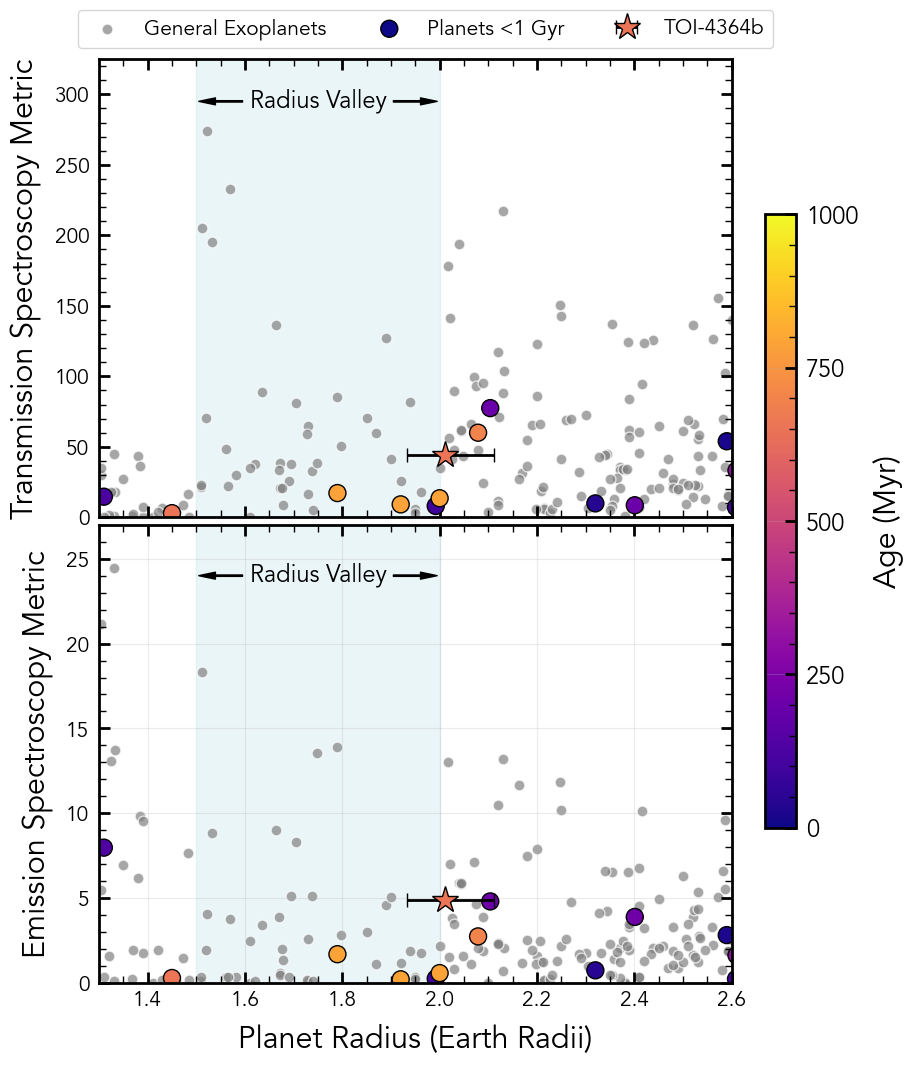}
    \caption{TSM (Top Panel) and ESM (Bottom Panel) values as a function of planet radius for \pname{} (star), the population of confirmed exoplanets (grey circles), and young planets with well-constrained ages ($<1$\,Gyr, larger circles). 
    The blue shaded region indicates the location of the radius valley as in \cite{fulton}.}
   \label{fig:TSM_and_ESM}
\end{figure}

The two panels in Figure~\ref{fig:TSM_and_ESM} illustrate the TSM and ESM values associated with \pname{} (calculations discussed in \S\ref{subsec:metrics}) compared to confirmed exoplanets with similar radii in the NASA Exoplanet Archive.\footnote{\url{https://exoplanetarchive.ipac.caltech.edu/}} 
We only illustrate confirmed transiting exoplanets with reliable radii. 

With reasonably strong TSM and ESM values, particularly among the population of young planets with well-constrained ages, \pname{} is a compelling candidate for detailed atmospheric characterization.
This system has the potential to offer insights into atmospheric retention and planet formation theories in young planetary systems. 

Noting its position at the upper edge of the radius valley, this planet could serve as a benchmark to test theories of atmospheric escape, investigating the relevant processes and timescales driving atmospheric loss. 
Valuable avenues to explore hydrodynamic escape and atmospheric mass loss include transit investigations of atomic hydrogen (Ly$\alpha$) \citep{2003Natur.422..143V} and metastable helium ($\lambda=10,830$\,\AA{}) absorption.

\section{Summary}\label{sec:summary}
In this study, we identified and characterized \pname{} --- a transiting mini-Neptune orbiting an M dwarf member of the Hyades cluster. This young planet is located on the upper edge of the radius valley. With a measured radius of \prad{} and a relatively short orbital period, the \pname{} likely has an inflated atmosphere due to its young age and high internal entropy. 

The discovery of \pname{} adds to our growing census of planetary systems orbiting young M dwarfs. As a cluster member, this newly-detected world benefits from a well-constrained stellar environment, enabling precise age estimates and meaningful comparisons with other exoplanets in similar stages of evolution. This context is crucial for understanding how planetary properties change over time under consistent stellar conditions.

\pname{}'s position near the upper edge of the radius valley offers a valuable case for studying the transition between gas-rich mini-Neptunes and rocky planets. 
Follow-up RV measurements are essential for determining the planet's mass, which will help to distinguish between a gas-rich envelope and a rocky, core-dominated planet, shedding light on the processes that shape planets in this size range.

While the TSM and ESM values for this system are modest when compared to the best candidates, they are high when compared to the population of young confirmed exoplanets. 

Further, the location of this M dwarf companion at the edge of the radius valley makes it a particularly compelling candidate for atmospheric characterization.
\pname{} is well-positioned for spectroscopic and outgassing observations, offering the potential to study its atmospheric composition and evolution.
In conclusion, \pname{} represents an important addition to the study of exoplanetary systems orbiting young M dwarfs. The system characteristics offer a promising opportunity to advance our understanding of planetary atmospheres and evolutionary pathways. 

\acknowledgments
We are thankful for our helpful conversations with Anne Dattilo.
MSF gratefully acknowledges the generous support provided by NASA through Hubble Fellowship grant HST-HF2-51493.001-A awarded by the Space Telescope Science Institute, which is operated by the Association of Universities for Research in Astronomy, Inc., for NASA, under the contract NAS 5-26555.
AWM was supported by grants from the NSF CAREER program (AST-2143763) and NASA's exoplanet research program (XRP 80NSSC21K0393).
This paper includes data collected by the \tess{} mission, which are publicly available from the Mikulski Archive for Space Telescopes (MAST). Funding for the \tess{} mission is provided by NASA’s Science Mission Directorate. All the \tess{} data used in this paper can be found in MAST: \dataset[10.17909/e64q-zy94]{http://dx.doi.org/10.17909/e64q-zy94}.
This research has made use of the Exoplanet Follow-up Observation Program (ExoFOP; DOI: 10.26134/ExoFOP5) website, which is operated by the California Institute of Technology, under contract with the National Aeronautics and Space Administration under the Exoplanet Exploration Program.
Some of the Observations in the paper made use of the High-Resolution Imaging instruments ‘Alopeke and Zorro. ‘Alopeke and Zorro were funded by the NASA Exoplanet Exploration Program and built at the NASA Ames Research Center by Steve B. Howell, Nic Scott, Elliott P. Horch, and Emmett Quigley. ‘Alopeke was mounted on the Gemini North, telescope Zorro was mounted on Gemini South telescope, both a part of the international Gemini Observatory, a program of NSF NOIRLab, which is managed by the Association of Universities for Research in Astronomy (AURA) under a cooperative agreement with the U.S. National Science Foundation on behalf of the Gemini Observatory partnership: the U.S. National Science Foundation (United States), National Research Council (Canada), Agencia Nacional de Investigaci\'{o}n y Desarrollo (Chile), Ministerio de Ciencia, Tecnolog\'{i}a e Innovaci\'{o}n (Argentina), Minist\'{e}rio da Ci\^{e}ncia, Tecnologia, Inova\c{c}\~{o}es e Comunica\c{c}\~{o}es (Brazil), and Korea Astronomy and Space Science Institute (Republic of Korea).
This work has made use of data from the European Space Agency (ESA) mission \emph{Gaia},\footnote{\url{https://www.cosmos.esa.int/gaia}} processed by the \emph{Gaia} Data Processing and Analysis Consortium (DPAC).\footnote{\url{https://www.cosmos.esa.int/web/gaia/dpac/consortium}} 
This research has made use of the VizieR catalogue access tool, CDS, Strasbourg, France. The original description of the VizieR service was published in A\&AS 143, 23. 
We acknowledge the use of public TOI Release data from pipelines at the \tess{} Science Office and the \tess{} Science Processing Operations Center. 
Resources supporting this work were provided by the NASA High-End Computing (HEC) Program through the NASA Advanced Supercomputing (NAS) Division at Ames Research Center for the production of the SPOC data products.

This publication makes use of data products from the Two Micron All Sky Survey, which is a joint project of the University of Massachusetts and the Infrared Processing and Analysis Center/California Institute of Technology, funded by the National Aeronautics and Space Administration and the National Science Foundation.
This work makes use of observations from the LCOGT network. Part of the LCOGT telescope time was granted by NOIRLab through the Mid-Scale Innovations Program (MSIP). MSIP is funded by NSF.
Based in part on observations obtained at the Southern Astrophysical Research (SOAR) telescope, which is a joint project of the Minist\'{e}rio da Ci\^{e}ncia, Tecnologia e Inova\c{c}\~{o}es (MCTI/LNA) do Brasil, the US National Science Foundation’s NOIRLab, the University of North Carolina at Chapel Hill (UNC), and Michigan State University (MSU).
Funding for the \tess{} mission is provided by NASA's Science Mission Directorate. 
%% K Barkaoui
The postdoctoral fellowship of KB is funded by F.R.S.-FNRS grant T.0109.20 and by the Francqui Foundation.
KAC and SNQ acknowledge support from the \tess{} mission via subaward s3449 from MIT. 

\facilities{\tess, \Gaia{} \citep{GaiaDR3}; Mikulski Archive for Space Telescopes \citep{MAST}; Wide-field Infrared Survey Explorer (\emph{WISE}); LCOGT 1m (Sinistro); LCOGT 1m (NRES);  Gemini: Gillett, South; Very Large Telescope; Southern Astrophysical Research (SOAR) Telescope}

\software{\texttt{AstroImageJ} \citep{Collins2017}, \texttt{astroquery} \citep{astroquery}, 
BANZAI \citep{banzai}, 
\texttt{batman} \citep{batman}, 
\texttt{dustmaps} \citep{dustmaps}, \texttt{edmcmc} \citep{vanderburgedmcmc},
\texttt{EXOFASTv2} \citep{Eastman2013}, \texttt{Lightkurve} \citep{lightkurve}, \texttt{matplotlib} \citep{matplotlib}, \texttt{PAdova and TRieste Stellar Evolution Code} \citep{parsec2012}, \texttt{PyAstronomy} \citep{pya}, \texttt{starrotate} \citep{starrotate},  \texttt{TRICERATOPS} \citep{Giacalone2021}, \texttt{TESSCUT} \citep{tesscut}}, \texttt{Vplanet} \citep{Barnes2020}

\pagebreak 
\bibliographystyle{aasjournalmod}
\bibliography{bibliography.bib}

\end{document}